\documentclass[journal,12pt,draftclsnofoot,onecolumn,twoside]{IEEEtran}

\usepackage{ifpdf}
 \ifpdf
     \usepackage[pdftex]{graphicx}
 \else
     \usepackage[dvips]{graphicx}
 \fi

\graphicspath{{figures/}}

\usepackage{cite}

\usepackage{epstopdf}					
\usepackage[cmex10]{amsmath}
\usepackage{amssymb,amsfonts}
\usepackage{algorithm}
\usepackage{algpseudocode}

\usepackage{mdwmath}
\usepackage{mdwtab}
\begin{document}
%
\title{A Novel Algorithm for Joint Bit and Power Loading for OFDM Systems with Unknown Interference}
%
%
%
\author{{Ebrahim Bedeer, 
Mohamed F. Marey, 
Octavia A. Dobre, 
Mohamed H. Ahmed, and 
Kareem E. Baddour \IEEEauthorrefmark{2}}\\
\IEEEauthorblockA{Faculty of Engineering and Applied Science, Memorial University of Newfoundland,
St. John's, NL, Canada\\ 
\IEEEauthorrefmark{2} Communications Research Centre, Ottawa, ON, Canada\\
Email: \{e.bedeer, mmarey, odobre, mhahmed\}@mun.ca, kareem.baddour@crc.ca}
}
\maketitle

\begin{abstract}
In this paper, a novel low complexity bit and power loading algorithm is formulated for orthogonal frequency division multiplexing (OFDM) systems operating in fading environments and in the presence of unknown interference. The proposed non-iterative algorithm jointly maximizes the throughput and minimizes the transmitted power, while guaranteeing a target bit error rate (BER) per subcarrier. Closed-form expressions are derived for the optimal bit and power distributions per subcarrier. The performance of the proposed algorithm is investigated through extensive simulations. A performance comparison with the algorithm in \cite{wyglinski2005bit} shows the superiority of the proposed algorithm with reduced computational effort.
%
\end{abstract}

\begin{IEEEkeywords}
Adaptive modulation, bit loading, frequency selective channels, joint optimization, OFDM, power loading, unknown interference.
\end{IEEEkeywords}

%

\section{Introduction}

Orthogonal frequency division multiplexing (OFDM) modulation is recognized as a robust and efficient transmission technique, as evidenced by its consideration for diverse communication systems and adoption by several wireless standards \cite{popescu2007narrowband, wang2011new, fazel2008multi}. The performance of OFDM systems can be significantly improved by dynamically adapting the transmission parameters, such as power, constellation size, symbol rate, coding rate/scheme, or any combination of these, according to the channel conditions or the wireless standard specifications \cite{hughes1988ensemble, de1998optimal, levin2001complete, wyglinski2005bit, fox1966discrete, song2002joint, sonalkar2000efficient, papandreou2005new, liu2009adaptive, goldfeld2002minimum, krongold2000computationally}. 

Generally speaking, the problem of optimally loading bits and power per subcarrier can be categorized into two main classes: \textit{rate maximization} (RM) and \textit{margin maximization} (MM). For the former, the objective is to maximize the achievable data rate, while for the latter the objective is to maximize the achievable system margin (i.e., minimizing the total transmit power given a target data rate) \cite{wyglinski2005bit}. Most of the algorithms for solving the RM and MM problems are variant of two main types: greedy algorithms \cite{hughes1988ensemble, de1998optimal, levin2001complete, fox1966discrete, sonalkar2000efficient, wyglinski2005bit, papandreou2005new, song2002joint} and water-filling based algorithms \cite{ goldfeld2002minimum, liu2009adaptive, krongold2000computationally}.

Greedy algorithms provide near optimal allocation, by incrementally allocating an integer number of bits, at the cost of high complexity. This type of algorithms was suggested by Hughes-Hartog in \cite{hughes1988ensemble}, where bits are successively allocated to subcarriers requiring the minimum incremental power until either the total transmit power exceeds the maximum power or the target BER rate is reached. Unfortunately, the algorithm is very complex and converges very slowly. Campello de Souza \cite{de1998optimal} and Levin \cite{levin2001complete} developed a complete and mathematically verifiable algorithm known as ``Levin-Campello" that significantly improves the work of Hughes-Hartog with lower complexity. Wyglinski \textit{et al.} \cite{wyglinski2005bit} proposed an incremental bit loading algorithm with uniform power in order to maximize the throughput while guaranteeing a target mean BER. This algorithm achieves nearly the optimal solution given in \cite{fox1966discrete} but with lower complexity. In \cite{song2002joint}, Song \textit{et~al.} proposed an iterative joint bit loading and power allocation algorithm based on \textit{statistical} channel conditions to meet a target BER, i.e., the algorithm loads bits and power per subcarrier based on long-term frequency domain channel conditions, rather than \textit{instantaneous} channel conditions as in \cite{hughes1988ensemble, de1998optimal, levin2001complete, wyglinski2005bit, fox1966discrete, sonalkar2000efficient, liu2009adaptive, goldfeld2002minimum, papandreou2005new, krongold2000computationally}. 

On the other hand, water-filling based algorithms formulate the RM and MM problems as constrained optimization problems that can be solved by classical optimization methods. The water-filling based algorithms allocate more power to subcarriers with higher instantaneous signal-to-noise ratio (SNR) (i.e., better channels) to maximize the throughput or minimize the BER on each subcarrier \cite{cover2004elements}. Typically, water-filling based algorithms allocate non-integer number of bits per each subcarrier; hence, it is generally followed by a rounding-off step to allocate an integer number of bits to the transmitted symbols across all subcarriers, which compromises performance for lower complexity. Liu and Tang \cite{liu2009adaptive} proposed a low complexity power loading algorithm with uniform bit loading that aims to minimize the transmit power while guaranteeing a target average BER. In \cite{goldfeld2002minimum}, Goldfeld \textit{et al.} proposed a quasi-optimal power loading algorithm that requires no iterations in order to minimize the overall BER with fixed constellation size across all subcarriers. 

When compared with previous works, in this paper we propose a low complexity, non-iterative algorithm that jointly maximizes the OFDM throughput and minimizes its total transmit power, subject to a constraint on the BER per subcarrier in the presence of unknown interference. Closed-form expressions for the optimal bit and power distributions are given. The performance of the proposed algorithm is investigated through extensive simulations, which also show that this approach outperforms Wyglinski's algorithm  presented in \cite{wyglinski2005bit} with reduced computational effort.

The remainder of the paper is organized as follows. Section \ref{sec:opt} presents the proposed joint bit and power loading algorithm. Simulation results are presented in Section \ref{sec:sim}, while conclusions are drawn in Section \ref{sec:conc}. 

\section{Proposed Algorithm} \label{sec:opt}
\subsection{Optimization Problem Formulation}
An OFDM system decomposes the signal bandwidth into a set of $N$ orthogonal narrowband subcarriers of equal bandwidth. Each subcarrier $i$ transmits $b_i$ bits using power $\mathcal{P}_i$, $i = 1, ..., N$. An unknown interference is assumed to affect the OFDM signal,  with a Gaussian distribution with zero mean and variance $\sigma^2_{u,i}$ per subcarrier $i$ \cite{bansal2008optimal,zhao2010power,sanguinetti2010frame,wu2005narrowband}; according to the central limit theorem \cite{papoulis1965probability}, such an assumption is valid assuming the interference comes from several independent sources.  A delay- and error-free feedback channel is assumed to exist between the transmitter and receiver for reporting channel state information.

In order to minimize the total transmit power and maximize the throughput subject to a BER constraint, the optimization problem is  formulated as 
\setlength{\arraycolsep}{0.0em}
\begin{equation}
\underset{\mathcal{P}_i}{\textup{Minimize}} \; \mathcal{P}_T=\sum_{i = 1}^{N}\mathcal{P}_i \quad \textup{and} \quad  \underset{b_i}{\textup{Maximize}} \; b_T=\sum_{i = 1}^{N}b_i, \nonumber
\end{equation}
\begin{equation}
\textup{subject to}  \quad  \textup{BER}_i \leq \textup{BER}_{th,i}, \: \: i = 1, ..., N, \label{eq:eq_1}
\end{equation}
where $\mathcal{P}_T$ and $b_T$ are the total transmit power and throughput, respectively, and $\textup{BER}_i$ and $\textup{BER}_{th,i}$ are the BER and threshold value of BER per subcarrier $i$, $i$ = 1, ..., $N$, respectively. An approximate expression for the BER per subcarrier $i$ in the case of $M$-ary QAM is given by\footnote[1]{This expression is tight within 1 dB for BER $\leq$ $10^{-3}$ \cite{chung2001degrees}.} \cite{liu2009adaptive,chung2001degrees}
\setlength{\arraycolsep}{0.0em}
\begin{eqnarray}
\textup{BER}_i &{} \approx  {}& 0.2 \: \textup{exp}\left ( -1.6 \frac{\mathcal{P}_i}{(2^{b_i} - 1)} \frac{\left | \mathcal{H}_i \right |^2}{(\sigma^2_n + \sigma^2_{u,i})} \right ), \label{eq:BER}
\end{eqnarray}
where $ \mathcal{H}_i $ is the channel gain of subcarrier $i$ and $\sigma^2_n$ is the variance of the additive white Gaussian noise (AWGN).

The multi-objective optimization function in (\ref{eq:eq_1}) can be rewritten as a linear combination of multiple objective function as follows
\setlength{\arraycolsep}{0.0em}
\begin{equation}
\underset{\mathcal{P}_i , b_i}{\textup{Minimize}}  \quad  \mathcal{F}(\textbf{\textit{P}},\textbf{\textit{b}}) = \: \left\{\alpha \sum_{i = 1}^{N}\mathcal{P}_i - (1-\alpha)\sum_{i = 1}^{N}b_i\right\}, \label{eq:p1} \nonumber 
\end{equation} \vspace{-10pt}
\begin{eqnarray}
\textup{subject to} \: \: \: g_i(\mathcal{P}_i,b_i) &{} = {}& 0.2 \: \textup{exp}\left ( \frac{- 1.6 \: \mathcal{C}_i \mathcal{P}_i}{2^{b_i} - 1} \right ) \nonumber\\  &{}  {}& - \textup{BER}_{th,i} \leq 0,  \hfill i = 1, ..., N, \label{eq:ineq_const}
\end{eqnarray}
where $\alpha$ ($0 < \alpha < 1$) is a constant whose value indicates the relative importance of one objective function relative to the other, $\mathcal{C}_i = \: \frac{\left | \mathcal{H}_i \right |^2}{\sigma^2_n +\sigma^2_{u,i}}$ is the channel-to-interference-plus-noise ratio for subcarrier $i$, and $\textbf{\textit{P}} = [\mathcal{P}_1, ..., \mathcal{P}_N]^T$ and $\textbf{\textit{b}} = [b_1, ..., b_N]^T$ are the \textit{N}-dimensional power and bit distribution vectors, respectively, with $[.]^T$ denoting the transpose operation.

\subsection{Optimal Bit and Power Distributions}
The problem in (\ref{eq:ineq_const}) can be solved by applying the method of Lagrange multipliers. Accordingly,  the inequality constraints are transformed to equality constraints by adding non-negative slack variables, $\mathcal{Y}_i^2$, $i$ = 1, ..., $N$ \cite{griva2009linear}. Hence, the constraints are rewritten as
\setlength{\arraycolsep}{0.0em}
\begin{eqnarray}
\mathcal{G}_i(\mathcal{P}_i,b_i,\mathcal{Y}_i) &{} = {}& g_i(\mathcal{P}_i,b_i) + \mathcal{Y}_i^2 = 0, \quad i = 1, ..., N, 
\label{eq:eq_const}
\end{eqnarray}
and the Lagrange function $\mathcal{L}$ is then expressed as
\setlength{\arraycolsep}{0.0em}
\begin{eqnarray}
\mathcal{L}(\textbf{\textit{P}},\textbf{\textit{b}},\textbf{\textit{Y}},\mathbf{\Lambda}) &{} = {}& \mathcal{F}(\textbf{\textit{P}},\textbf{\textit{b}}) + \sum_{i = 1}^{N} \lambda_i \: \mathcal{G}(\mathcal{P}_i,b_i,\mathcal{Y}_i), \nonumber \\
 &{} = {}& \alpha \sum_{i = 1}^{N}\mathcal{P}_i - (1-\alpha)\sum_{i = 1}^{N}b_i \nonumber \\ 
& & + \sum_{i = 1}^{N} \lambda_i\:\Bigg[ 0.2 \: \textup{exp}\left ( \frac{- 1.6 \: \mathcal{C}_i \mathcal{P}_i}{2^{b_i} - 1} \right ) - \textup{BER}_{th,i} \nonumber \\ 
& & \hspace{4.5cm}+ \mathcal{Y}_i^2\Bigg],
\end{eqnarray}
where $\mathbf{\Lambda} = [\lambda_1, ..., \lambda_N]^T$ and $\textbf{\textit{Y}} = [\mathcal{Y}_1^2, ..., \mathcal{Y}_N^2]^T$ are the vectors of Lagrange multipliers and slack variables, respectively. A stationary point can be found when $\nabla \mathcal{L}(\textbf{\textit{P}},\textbf{\textit{b}},\textbf{\textit{Y}},\mathbf{\Lambda}) = 0$ (where $\nabla$ denotes the gradient), which yields
\setlength{\arraycolsep}{0.0em}
\begin{eqnarray}
\frac{\partial \mathcal{L}}{\partial \mathcal{P}_i} &{} = {}& \alpha - 0.2 \: \lambda_i \frac{1.6 \: \mathcal{C}_i}{2^{b_i}-1} \: \textup{exp}\left ( \frac{- 1.6 \: \mathcal{C}_i \mathcal{P}_i}{2^{b_i} - 1} \right ) = 0,\label{eq:eq1}\\
\frac{\partial \mathcal{L}}{\partial b_i} &{} = {}& -(1 - \alpha) + 0.2 \ln (2) \: \lambda_i \frac{1.6 \: \mathcal{C}_i \mathcal{P}_i 2^{b_i}}{(2^{b_i}-1)^2} \: \nonumber \\ & & \hspace{2.5cm} \times \: \textup{exp}\left ( \frac{- 1.6 \: \mathcal{C}_i \mathcal{P}_i}{2^{b_i} - 1} \right ) = 0,\label{eq:eq2}\\
\frac{\partial \mathcal{L}}{\partial \lambda_i} & = & 0.2 \: \textup{exp}\left ( \frac{- 1.6 \: \mathcal{C}_i \mathcal{P}_i}{2^{b_i} - 1} \right ) - \textup{BER}_{th,i} + \mathcal{Y}_i^2= 0, \label{eq:eq3}\\
\frac{\partial \mathcal{L}}{\partial \mathcal{Y}_i} & = & 2\lambda_i \mathcal{Y}_i = 0. \label{eq:eq4}
\end{eqnarray}
It can be seen that (\ref{eq:eq1}) to (\ref{eq:eq4}) represent $4N$ equations in the $4N$ unknown components of the vectors $\textbf{\textit{P}}, \textbf{\textit{b}}, \textbf{\textit{Y}}$, and $\mathbf{\Lambda}$. By solving (\ref{eq:eq1}) to (\ref{eq:eq4}), one obtains the solution $\textbf{\textit{P}}^*, \textbf{\textit{b}}^*$, the slack variable vector $\textbf{\textit{Y}}^*$, and the Lagrange multiplier vector $\mathbf{\Lambda}^*$. Equation (\ref{eq:eq4}) implies that either $\lambda_i$ = 0 or $\mathcal{Y}_i$ = 0; hence, two possible solutions exist, and we are going to investigate each case independently.   

--- \textit{Case 1} ($\mathcal{Y}_i = 0$): 
From (\ref{eq:eq1}) and (\ref{eq:eq2}), we can relate $\mathcal{P}_i$ and $b_i$ as
\setlength{\arraycolsep}{0.0em}
\begin{eqnarray}
\mathcal{P}_i &{} = {}& \frac{1- \alpha}{\alpha \ln(2)}(1 - 2^{-b_i}), \label{eq:eq8}
\end{eqnarray}
with $\mathcal{P}_i \geq 0$ if and only if $b_i \geq 0$. By substituting (\ref{eq:eq8}) into (\ref{eq:eq3}), one obtains the solution
\setlength{\arraycolsep}{0.0em}
\begin{eqnarray}
b_i^* = \frac{1}{\log(2)}\log\Bigg(- \frac{1-\alpha }{\alpha \ln(2)} \frac{1.6 \: \mathcal{C}_i}{\ln(5 \: \textup{BER}_{th,i})}\Bigg). \label{eq:eq10}
\end{eqnarray}
Consequently, from (\ref{eq:eq8}) one gets
\setlength{\arraycolsep}{0.0em}
\begin{eqnarray}
\mathcal{P}_i^* = \frac{1-\alpha }{\alpha \ln(2)}\Bigg( 1 - \Big(- \frac{1-\alpha }{\alpha \ln(2)} \frac{1.6 \: \mathcal{C}_i}{\ln(5 \: \textup{BER}_{th,i})} \Big)^{-1} \Bigg). \label{eq:eq12}
\end{eqnarray}
Since (\ref{eq:BER}) is valid for $M$-ary QAM, $b_i$ should be greater than 2. From (\ref{eq:eq10}), to have $b_i \geq 2$, the channel-to-noise ratio per subcarrier, $\mathcal{C}_i$, must satisfy the condition
\setlength{\arraycolsep}{0.0em}
\begin{eqnarray}
\mathcal{C}_i \geq - \frac{4}{1.6} \: \frac{\alpha \ln(2)}{1-\alpha} \: \ln(5\:\textup{BER}_{th,i}), \qquad i = 1, ..., N. \label{eq:condition}
\end{eqnarray}

--- \textit{Case 2} ($\lambda_i = 0$): 
By following a similar procedure as in Case 1, one can show that $\nabla \mathcal{L}(\textbf{\textit{P}},\textbf{\textit{b}},\textbf{\textit{Y}},\mathbf{\Lambda}) = 0$ results in an underdetermined system of $N$ equations in $4N$ unknowns, and, hence, no unique solution can be reached.

The obtained solution represents a minimum of $\mathcal{F}(\textbf{\textit{P}},\textbf{\textit{b}})$ if the Karush-Kuhn-Tucker conditions are satisfied \cite{kuhn1951nonlinear}. Given that our stationary point ($b_i^*$, $\mathcal{P}_i^*$) in (\ref{eq:eq10}) and (\ref{eq:eq12}) exists at $\mathcal{Y}_i = 0$, $i$ = 1, ..., $N$, the Karush-Kuhn-Tucker conditions can be written as 
\setlength{\arraycolsep}{0.0em}
\begin{eqnarray}
\frac{\partial \mathcal{F}}{\partial \mathcal{P}_i} + \sum_{j = 1}^{N}\lambda_j \: \frac{\partial g_j}{\partial \mathcal{P}_i} &{} = {}& 0, \label{eq:KH1}\\
\frac{\partial \mathcal{F}}{\partial b_i} +  \sum_{j = 1}^{N}\lambda_j \: \frac{\partial g_j}{\partial b_i} &{} = {}& 0, \label{eq:KH2}\\
\lambda_j &{} > {}& 0, \label{eq:KH3}
\end{eqnarray}
$i$, $j$ = 1, ..., $N$. One can easily prove that these conditions are fulfilled, as follows.

\textit{Proof of} (\ref{eq:KH1})-(\ref{eq:KH3}): From (\ref{eq:eq1}), one finds
\setlength{\arraycolsep}{0.0em}
\begin{eqnarray}
\lambda_j = \alpha \Bigg[ 0.2 \: \frac{1.6 \: C_j}{2^{b_j}-1} \: \textup{exp}\Big( \frac{-1.6 \: C_j P_j}{2^{b_j}-1}  \Big)  \Bigg]^{-1}, \label{eq:lambda}
\end{eqnarray}
which is positive for all values of $j$, and hence it satisfies (\ref{eq:KH3}). Furthermore, by substituting (\ref{eq:eq10}), (\ref{eq:eq12}), and (\ref{eq:lambda}) into (\ref{eq:KH1}) and (\ref{eq:KH2}), one can easily verify that the rest of the Karush-Kuhn-Tucker conditions are satisfied, and, thus, the solution ($\textbf{\textit{b}}^*,\textbf{\textit{P}}^*$) represents an optimum point. \hfill$\blacksquare$ 

\subsection{Proposed Joint Bit and Power Loading Algorithm} 
The idea behind the proposed algorithm is to check the channel-to-interference-plus-noise ratio per subcarrier, $\mathcal{C}_i$, against the condition in (\ref{eq:condition}). If this is fulfilled, the optimal bit and power is given by (\ref{eq:eq10}) and (\ref{eq:eq12}), respectively; otherwise, the corresponding subcarrier is nulled. The final bit and power allocation is reached after rounding the non-integer number of bits obtained from (\ref{eq:eq10}) to the \textit{nearest} integer and recalculating the allocated power according to (\ref{eq:BER}).  The proposed algorithm can be formally stated as follows. 

\floatname{algorithm}{}
\begin{algorithm}
\renewcommand{\thealgorithm}{} 
\caption{\textbf{Proposed Algorithm}} 
\begin{algorithmic}[1]
\State \textbf{INPUT} The AWGN variance $\sigma^2_n$, channel gain per subcarrier $i$ ($\mathcal{H}_i$), target BER per subcarrier $i$ ($\textup{BER}_{th,i}$), and weighting factor $\alpha$.
        \algstore{myalg}
  \end{algorithmic}
\end{algorithm}

\floatname{algorithm}{}
\begin{algorithm}
 \renewcommand{\thealgorithm}{} 
  \caption{\textbf{Proposed Algorithm} (continued)}
  \begin{algorithmic}
      \algrestore{myalg}
\For{$i$ = 1, ..., $N$}    
\If{$\mathcal{C}_i \geq - \frac{4}{1.6} \: \frac{\alpha \ln(2)}{1-\alpha} \: \ln(5\:\textup{BER}_{th,i})$}
\State - $b_i^*$ and $\mathcal{P}_i^*$ are given by (\ref{eq:eq10}) and (\ref{eq:eq12}), respectively.
\If{$b_i^* \geq$ 2}
\State - Round $b_i^*$ to the nearest integer.
\State - Recalculate $\mathcal{P}_i^*$ according to (\ref{eq:BER}).
\Else
\State - Null the corresponding subcarrier $i$.
\EndIf
\Else
\State - Null the corresponding subcarrier $i$.
\EndIf
\EndFor
\State \textbf{OUTPUT} $b_i^*$ and $\mathcal{P}_i^*$, $i$ = 1, ..., $N$.
\end{algorithmic} 
\end{algorithm}

%
\section{Numerical Results} \label{sec:sim}

This section investigates the performance of the proposed algorithm in terms of the achieved average throughput and average transmit power in the presence of unknown interference, along with the algorithm computational complexity. Furthermore, the performance and complexity of the proposed algorithm are compared with that of the algorithm presented in \cite{wyglinski2005bit}.

\subsection{Simulation Setup}
An OFDM system with a total of $N$ = 128 subcarriers is considered. Without loss of generality, the unknown interference is assumed to affect $N_u$ subcarriers, $N_u$ = 40, with exponentially distributed variance across the affected subcarriers\footnote[2]{Such distribution is chosen to approximate the effect of narrowband interference signals on the OFDM subcarriers noticed from simulations; however other distributions can be straightforwardly applied.}, i.e., $\sigma^2_{u,i} = e^{-\beta x}$, $\beta = - 0.25$, $x$ = 0,...,$N_u-1$. For simplicity, the BER constraint per subcarrier, $\textup{BER}_{th,i}$, is assumed to be the same for all subcarriers and set to $10^{-4}$. The channel impulse response $h(n)$ of length $N_{ch}$ is modeled as independent complex Gaussian random variables with zero mean and exponential power delay profile \cite{marey2007analysis}
\setlength{\arraycolsep}{0.0em}
\begin{eqnarray}
\mathbb{E}\{\left | h(n) \right |^2\} = \sigma_h^2 \: e^{-n\Xi}, \qquad n = 0, 1, ..., N_{ch}-1,
\end{eqnarray}
where $\sigma_h^2$ is a constant chosen such that the average energy per subcarrier is normalized to unity, i.e., $\mathbb{E}\{\left | \mathcal{H}_i \right |^2\}$ = 1, and $\Xi$ represents the decay factor. Representative results are presented in this section and were obtained by repeating Monte Carlo trials for $10^{5}$ channel realizations with a channel length $N_{ch} = 5$ taps and decay factor  $\Xi = \frac{1}{5}$.

\subsection{Performance of the Proposed Algorithm}
\begin{figure*}[!t]
	\centering
		\includegraphics[width=1.00\textwidth]{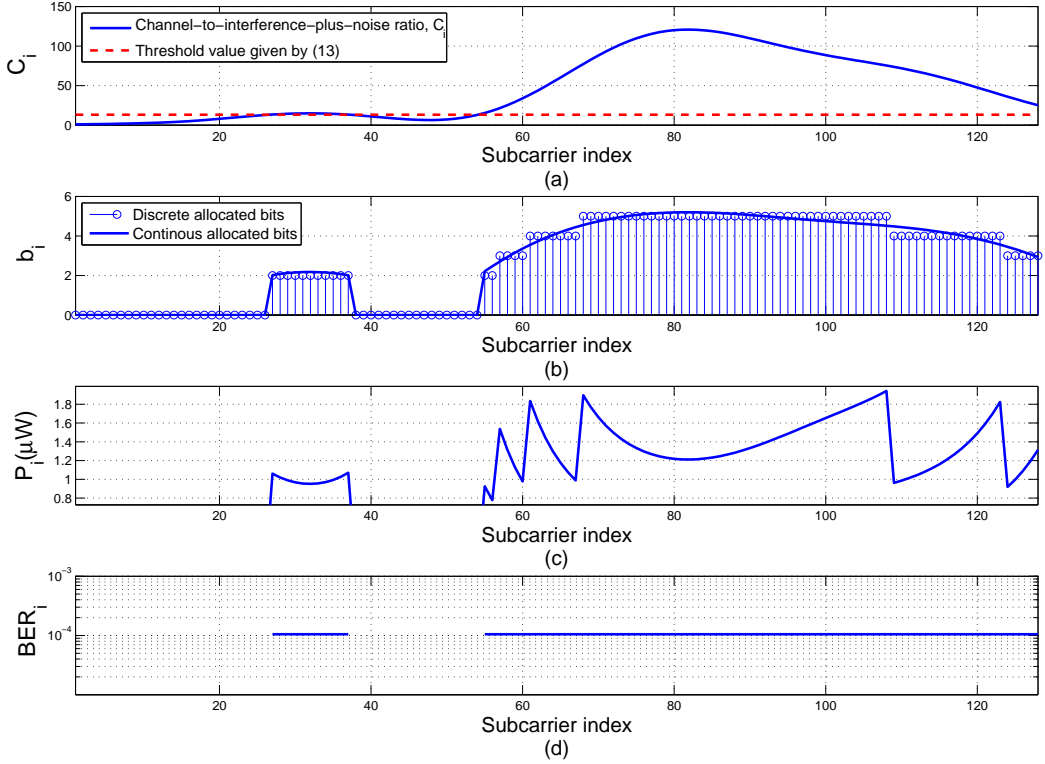}
	\caption{Example of allocated bits and power per subcarrier for a typical channel realization with $\sigma_n^2 = 10^{-3}$ $\mu W$, $N$ = 128, $\alpha$  = 0.5, $\textup{BER}_{th,i}$ = $10^{-4}$, and $N_u$ = 40.}
	\label{fig:ch_realization}
\end{figure*}

Fig. \ref{fig:ch_realization} illustrates the allocated bits and power computed using (\ref{eq:eq10}) and (\ref{eq:eq12}), respectively, for an example channel realization with $\sigma_n^2 = 10^{-3}$ $\mu W$ and $\alpha$  = 0.5. It can be seen from the plots in Fig.~\ref{fig:ch_realization} that when the channel-to-interference-plus-noise ratio per subcarrier, $\mathcal{C}_i$, exceeds the value in (\ref{eq:condition}), the number of bits and power allocated per subcarrier are non-zero. As expected, (\ref{eq:eq10}) yields a non-integer number of allocated bits per subcarrier, which is not suitable for practical implementations. This value is rounded to the nearest integer, as shown in Fig.~\ref{fig:ch_realization} (b), and the modified value of the allocated power per subcarrier to maintain the same $\textup{BER}_{th,i}$ is determined using (\ref{eq:BER}).  

\begin{figure}[!t]
	\centering
		\includegraphics[width=0.50\textwidth]{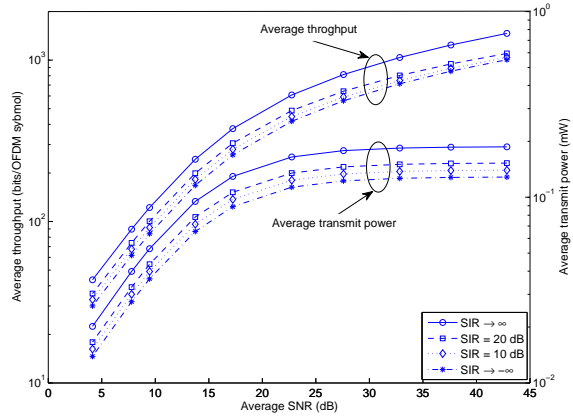}
	\caption{Effect of the average SNR and SIR on the average throughput and average transmit power when $\alpha$  = 0.5 and $N_u$ = 40.}
	\label{fig:inter_all_N_128_final_SIR}
\end{figure}

Fig. \ref{fig:inter_all_N_128_final_SIR} depicts the average throughput and average transmit power when $\alpha$ = 0.5 as a function of average SNR, for different average SIR values\footnote[3]{The average SNR is calculated by averaging the instantaneous SNR values per subcarrier over the total number of subcarriers and the total number of channel realizations, respectively. Moreover, the average SIR is calculated by averaging the instantaneous SIR values per subcarrier over the total number of affected subcarriers and the total number of channel realizations, respectively.}. For an average SIR $\rightarrow \infty$, i.e., no interference, and average SNR $\leq$ 24 dB, one finds that both the average throughput and the average transmit power increase as the average SNR increases, while for an average SNR $\geq$ 24 dB, the transmit power saturates while the throughput continues to increase. This observation can be explained as follows. For lower values of average SNR, the corresponding values of $\mathcal{C}_i$ result in the nulling of many subcarriers in (\ref{eq:condition}). By increasing the average SNR, the number of used subcarriers increases, resulting in a noticeable increase in the throughput and power. Apparently, for SNR $\geq$ 24 dB, all subcarriers are used and our proposed algorithm essentially minimizes the average transmit power by keeping it constant, while increasing the average throughput. On the other hand, reducing the average SIR, increases the effect of the interference on the OFDM system and more subcarriers are nulled, hence, both the average throughput and transmit power decrease as in Fig.~\ref{fig:inter_all_N_128_final_SIR}. For SIR $\rightarrow -\infty$, i.e., very strong interference, all $N_u$ affected subcarriers are nulled and both the average throughput and transmit power are affected accordingly. Note that this provides a lower performance bound for a given number of affected subcarriers, $N_u$. 

\begin{figure}[!t]
	\centering
		\includegraphics[width=0.50\textwidth]{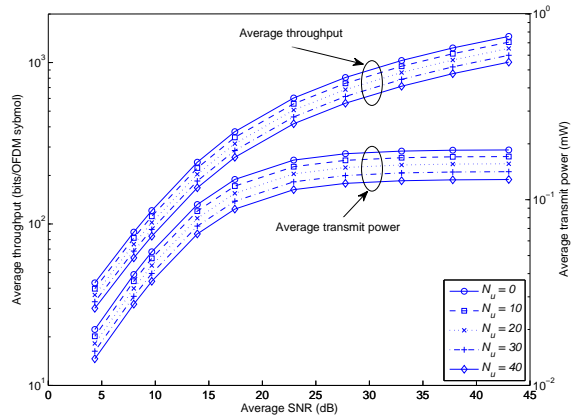}
	\caption{Effect of the average SNR and $N_u$ on the average throughput and average transmit power when $\alpha$  = 0.5.}
	\label{fig:inter_all_N_128_final_Nc}
\end{figure}

Fig. \ref{fig:inter_all_N_128_final_Nc} shows the average throughput and average transmit power as a function of average SNR, for diverse $N_u$ and $\alpha$  = 0.5. For $N_u$ = 0, i.e., no interference, the average throughput increases as the average SNR increases, while, the average transmit power increases for lower values of average SNR and saturates for higher values as discussed earlier. As $N_u$ increases, more subcarriers are affected by the interference, and hence, both the average throughput and average transmit power decreases. 

\begin{figure}
	\centering
		\includegraphics[width=0.50\textwidth]{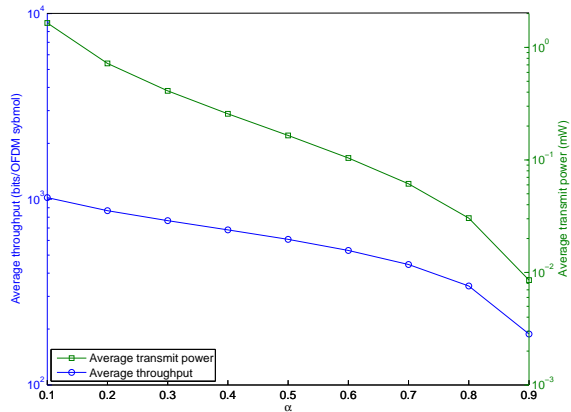}
	\caption{Average throughput and average transmit power as a function of weighting factor $\alpha$ for $\sigma_n^2 = 10^{-3}$ $\mu W$ and $\sigma^2_{u,i} = 0$.}
	\label{fig:proposed_alpha}
\end{figure}

In Fig. \ref{fig:proposed_alpha}, the average throughput and average transmit power are plotted as a function of the weighting factor $\alpha$, for $\sigma_n^2 = 10^{-3}$ $\mu$W and $\sigma_{u,i}^2$ = 0 (no interference). By increasing $\alpha$, more weight is given in our problem formulation to minimizing the transmit power over maximizing the throughput. In this case, the corresponding reduction in the minimum transmit power is accompanied by a reduction in the maximum throughput.

\begin{figure}
	\centering
		\includegraphics[width=0.50\textwidth]{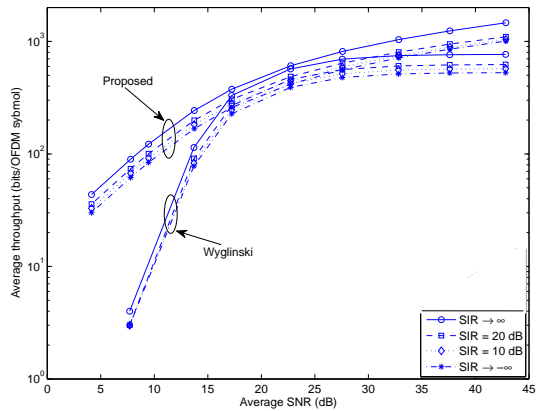}
	\caption{Average throughput as a function of average SNR and average SIR when $\alpha$ = 0.5, for the proposed algorithm and Wyglinski's algorithm in \cite{wyglinski2005bit}.}
	\label{fig:wyg_comp}
\end{figure}

In Fig. \ref{fig:wyg_comp}, the throughput achieved by the proposed algorithm is compared with that obtained by Wyglinski's algorithm presented in \cite{wyglinski2005bit} for the same operating conditions. To make a fair comparison, the uniform power allocation used by the allocation scheme in [7] is computed by dividing the average transmit power allocated by our algorithm by the total number of subcarriers. As can be seen in Fig.~\ref{fig:wyg_comp}, the proposed algorithm provides a significantly higher throughput than the scheme in \cite{wyglinski2005bit} for low average SNR values. This result demonstrates that optimal allocation of transmit power is crucial for low power budgets. Furthermore, for increasing average SNR values, the average transmit power is constant as seen in Fig.~\ref{fig:inter_all_N_128_final_SIR} for values $\geq$ 24 dB, which in turn results in a saturating throughput for Wyglinski's algorithm. In contrast, the proposed algorithm provides an increasing throughput for the same range of SNR values. As expected, increasing the effect of the interference, i.e., decreasing SIR, reduces the average throughput.

The improved performance of the proposed joint bit and power allocation algorithm does not come at the cost of additional complexity.  The proposed algorithm is non-iterative with a worst case computational complexity of $\mathcal{O(N)}$, while Wyglinski's algorithm is iterative with a worst case computational complexity of $\mathcal{O}(\mathcal{N}^2)$. 

\section{Conclusion} \label{sec:conc}
In this paper, we proposed a novel algorithm that jointly maximizes the throughput and minimizes the transmit power given a BER constraint per subcarrier for OFDM systems in the presence of unknown interference. Closed-form expressions for the optimal bit and power loading per subcarrier were derived. Simulation results demonstrated that the proposed algorithm outperforms the one presented in \cite{wyglinski2005bit} under the same operating conditions, while requiring reduced computational effort.


%

%

\section*{Acknowledgment}

This work has been supported in part by the Communications Research Centre, Canada.%
\end{document}